\documentclass[12pt,5p,twocolumn]{elsarticle}

\usepackage{amssymb}
\usepackage{amsmath}
\usepackage{lineno} 
\usepackage[usenames,dvipsnames]{xcolor}
\usepackage{float}
\usepackage{dblfloatfix}
\usepackage{babel}
\usepackage{upgreek}
\usepackage{ulem}
\usepackage{cuted}
\usepackage{caption}
 
 \graphicspath{{./}}
 \biboptions{numbers,sort&compress} 

\journal{Powder Technology}

\begin{document}

\begin{frontmatter}

\title{A cantilever-based method for probing adhesion force and kinetic detachment energy in cohesive granular materials}
\author{Maxime Lajeunesse$^{a}$}

\affiliation[label1]{organization={Aix Marseille University,  CNRS, IUSTI},
          city={Marseille},
          country={France}}

\author{Francisco M. Rocha$^{a,b}$} 
\affiliation[label2]{organization={Institut Franco-Argentin de Dynamique des Fluides pour l'Environnement (IFADyFE), CNRS IRL 2027, University of Buenos Aires, CONICET},
     city={Buenos Aires},
     country={Argentina}}
         
\author{Denis Dumont$^{a}$}          
          
\author{Vincent Bertin$^{c}$} 
\affiliation[label2]{organization={Universit\'e C\^ote d'Azur, CNRS, INPHYNI},
          city={Nice},
          country={France}}        
          
\author{Yoanh Moratille$^{d}$} 
          
\author{Nathalie Sintes-Zydowicz$^{d}$} 
\affiliation[label2]{organization={Universit\'e Lyon 1 Claude Bernard, CNRS, Ing\'enierie des Mat\'eriaux Polym\`eres, UMR 5223},
          city={Villeurbanne},
          country={France}}       
          
 \author{Eric Drockenmuller$^{d}$} 
          
\author{Olivier Pouliquen$^{a}$} 
\ead{olivier.pouliquen@univ-amu.fr}

\begin{abstract}
Interparticle adhesion plays a critical role in the flow behavior of cohesive granular materials, yet quantitative characterization at the particle scale remains challenging. Here, we present a cantilever-based method that combines a glass-beam cantilever for force measurements with a novel catapult mode for energy measurements, enabling the determination of both quasi-static detachment force and kinetic detachment energy. Applied to three different cohesive materials: polymer-coated silica particles, wet granular media, and cross-linked poly(2-ethylhexyl acrylate) (PEHA) microparticles, this method reveals a fundamental decoupling between force and energy, suggesting that both quantities are equally critical for characterizing grain properties.
\end{abstract}


\end{frontmatter}{}


\section{Introduction}

While substantial progress has been made in understanding and modeling dry granular materials~\cite{kamrin2024advances}, predicting the flow of cohesive granular materials, such as wet sand, snow, and powders, remains a significant scientific challenge~\cite{rognon2006rheophysics,badetti2018,vo2020additive,Pouliquen2025}. These materials are ubiquitous in industry, where issues like jamming, clogging, and flow localization pose serious challenges for process optimization and product manufacturing. Cohesive granular materials are also central to large-scale geophysical flows, such as debris flows and snow avalanches, which are among the most dangerous natural hazards, threatening both people and infrastructure. 

Adhesion between particles can arise from various physical interactions, including van der Waals forces, capillary bridges, or chain entanglement in the case of polymer-coated particles. A major challenge lies in linking these microscopic interactions to macroscopic flow behavior \cite{li2004,Castellanos05,lumay2012measuring}. A critical first step toward addressing this challenge is to accurately characterize interparticle interactions, enabling their control and providing insight into their connection to rheological properties~\cite{sharma2025experimental}. In this paper, we propose a simple yet powerful innovative experimental method to accurately measure both adhesion detachment forces and kinetic detachment energy between adhesive particles, a crucial milestone towards a better understanding of the flow behavior of cohesive granular materials.

Despite their importance, measurements of adhesive properties at the particle scale are scarce in powder technology. For sufficiently large particles ($d > 1$ mm), sensitive force sensors can be used to measure the force required to detach two particles in contact~\cite{pitois2000liquid,sharma2025experimental}. For smaller particles ($d<100$ $\mu$m), Atomic Force Microscopy (AFM) is an efficient alternative. AFM has been extensively developed and used in the suspensions and colloids community to accurately probe particle properties \cite{ducker1991,kappl2002,scherrer2024}. For dry particles and powders, the use of AFM is less common but has been employed in several studies \cite{jones2003,li2004,fuchs2014,colbert2015}. By attaching one particle to the AFM cantilever and another to a substrate, it is possible to probe adhesion forces. However, this technique is relatively tedious, limited to particles smaller than the typical size of AFM cantilevers ($d<100$ $ \mu$m), and does not easily allow for a large number of measurements to obtain meaningful statistics.

Other approaches have been developed for layers of grains rather than individual particles, enabling adhesion measurements between a surface and a large number of particles. One technique involves placing grain-covered plates in a centrifuge and observing grain detachment as the rotational speed gradually increases~\cite{booth1987,podczeck1995}. Another method consists of subjecting the plate to an impact and counting the number of grains detached \cite{wanka2013,deng2021,pedrolli2023}. While these methods provide estimates of the adhesion forces at play, they lack accuracy in characterizing interparticle interactions.

In this study, we present a new method to measure and characterize the adhesive properties of particles in the 100 $\mu$m - 1 mm range. We sought a more flexible approach than AFM, enabling large-scale statistical measurements. The method is largely inspired by techniques used in biophysics for probing cell responses to small forces, where a thin glass beam is used as a cantilever~\cite{israelachvili2011intermolecular,desprat2006microplates}. We demonstrate that this method allows for easy characterization of numerous particles and provides statistical insights. Additionally, we modified the setup to function as a catapult, measuring the minimum launching velocity necessary to detach one particle from another. This measurement resembles percussion experiments proposed in the literature \cite{deng2021,pedrolli2023} to estimate adhesion between a surface and particles but is adapted for single pairs of particles. This approach provides a measure of a kinetic detachment energy. While the literature discusses energy as a crucial parameter that may control the flowing regime of cohesive granular materials~\cite{RoyEtAl2016,mandal2020insights,mandal2021rheology,MacauleyRognon2021,lamarche2021}, such a quantity has yet to be thoroughly documented and studied. Our method introduces a direct experimental way to quantify it.

To demonstrate the flexibility of our method, we apply these techniques to analyze the adhesive properties of various model cohesive materials where adhesion stems from different origins. We analyze polymer-coated silica particles, a Cohesion-Controlled Granular Material (CCGM) recently proposed to create sticky particles where adhesion is controlled by the coating thickness \cite{gans2020cohesion}. We also probe wet particles, where small amounts of liquid are mixed within the medium, inducing the formation of capillary bridges between individual grains~\cite{Herminghaus2005,MitaraiNori2006,ScheelEtAl2008}. The last material used in this study consists of synthesized polymer particles~\cite{MoratileEtAl2022}, obtained through membrane emulsification, whose mechanical properties differ from the others by being much softer.

The paper is organized as follows. In Section~\ref{sec:particles}, we describe the adhesive particles used throughout the study. In Section~\ref{sec:force}, we detail the experimental setup designed to measure adhesion forces between cohesive grains and present the results for the CCGM particles. Section~\ref{sec:energy} describes how the setup was adapted to assess kinetic detachment energies. Finally, in Section~\ref{sec:DiffMaterials}, we demonstrate the flexibility of our method by applying it to other cohesive materials described above.

\begin{figure*}[t!]
  \centering
  \includegraphics[width=\textwidth ]{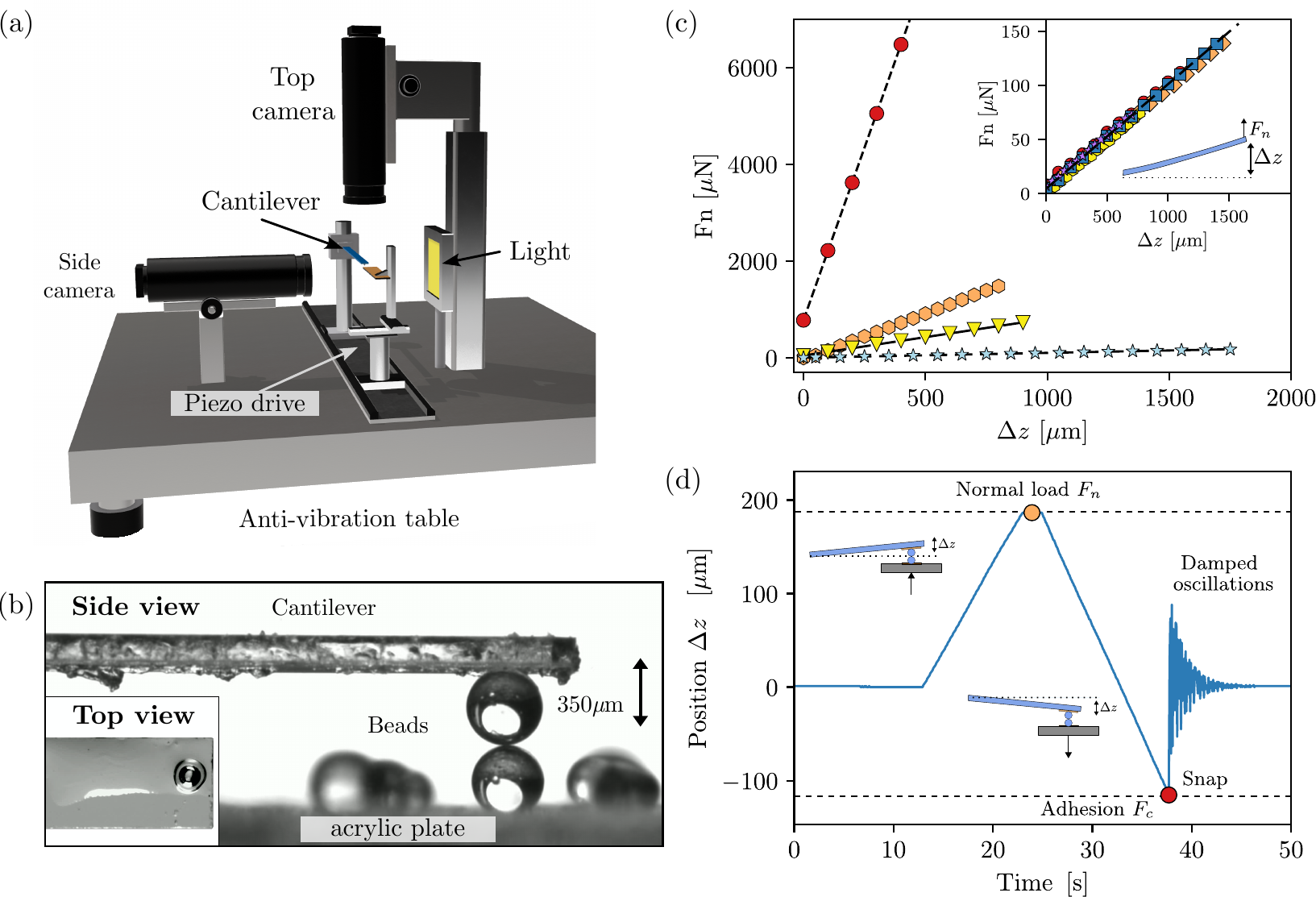}
  \caption{(a) Sketch of the experimental cantilever setup. (b) Side and top views of a typical field of view during force measurements. (c) Calibration of the force-deflection relationship for glass beams (width $w = 1$ mm) with varying thickness $e$ and length $L$; $e=400\,\mu$m and  $L=6$ cm (red circles); $e=400\,\mu$m and $L=8$  cm (orange diamonds); $e=145\,\mu$m and $L=6$ cm (yellow triangles); $e=145\,\mu$m and $L=8$ cm (blue stars). Dashed lines show predictions from Eq.~\eqref{eq:raideur}. Inset: Reproducibility test for four beams with identical geometry  ($e=145\,\mu$m and $L=8$ cm) providing the cantilever stiffness $k=0.095\pm 0.002$ N/m. (d) Time evolution of the vertical deflection $\Delta z$ measured by the side camera during one attachment-detachment cycle between two particles. }
  \label{setup}
\end{figure*}

\section{The adhesive particles}
\label{sec:particles}
Experiments were performed using three cohesive granular materials, each exhibiting a different origin of interparticle adhesion.

The first material consists of spherical glass particles coated with a thin layer of polyborosiloxane (PBS). For millimeter-sized particles, the adhesion force between two CCGM particles has been shown to depend on the average thickness $b$ of the PBS coating~\cite{gans2020cohesion}. Here, we extend this approach to silica particles with diameters $d = 199 \pm 16$ $\mu$m and $350 \pm 67$ $\mu$m, and average PBS coating thicknesses ranging from 0 to 140 nm, prepared according to the protocol described earlier~\cite{gans2020cohesion}. Particles are first mixed with boric acid dissolved in purified water heated at 60°C  and the PDMS. The mixture is  stirred for 90 minutes in a mixer at 110°C to evaporate the water. At the end of the process, the polymer cross-links with the silica particles and creates a quasi-homogeneous coating. We provide a more detailed explanation of the coating process in \ref{appendix0}.

The second material consists of the same glass beads with $d = 199 \pm 16 \, \mu$m mixed with a Newtonian, non-volatile liquid composed of 60 wt\% glycerol and 40 wt\% water. The medium was prepared by adding liquid corresponding to 0.1 wt\% of the particle mass and thoroughly mixing the system to obtain a wet granular medium in the pendular regime, where interparticle adhesion arises from liquid capillary bridges between pairs of particles~\cite{MitaraiNori2006,ScheelEtAl2008}.

The  third   material  consists of  cross-linked PEHA microparticles synthesized using a procedure adapted from earlier work ~\cite{MoratileEtAl2022}, with an averaged diameter $d = 232 \pm 34 $ $\mu$m. The microparticles are composed of tailor-made cross-linked poly(2-ethylhexyl acrylate) (PEHA) and are obtained through membrane emulsification-assisted suspension radical copolymerization. The synthesis of these polymer microparticles is fully detailed in \ref{appendix1}. Since the glass transition temperature of PEHA is below ambient temperature ($T_g = $ -59 °C), the particles are relatively soft. The Young's modulus, estimated from nanoindentation, is approximately $E=5.2 $ MPa, which is $10^4$ times smaller than that of the glass particles for which $E=50$ GPa.

\section{Measuring Adhesion Force}\label{sec:force} 

\subsection{Experimental Setup}

Classical methods for measuring adhesion forces often rely on cantilevers, i.e., elastic beams whose deformation provides a measure of the applied force. This principle underlies conventional Atomic Force Microscopy (AFM), in which cantilever deflection is accurately measured by tracking the displacement of a reflected laser beam. Our experimental setup, shown in Fig.~\ref{setup}a, is based on a cantilever design inspired by systems developed for probing cell mechanics~\cite{desprat2006microplates,wu2018cell}.

The cantilever consists of a thin glass beam with a length of 6-8 cm, width of 1 mm, and thickness of 145-400 $\mu$m. These dimensions can be adjusted to tune the cantilever stiffness to the expected force range. One end of the cantilever is fixed to a vertical support, while a grain is attached to the free end using a repositionable adhesive (3M RePositionable 75). The adhesive strength of the glue is significantly larger than the interparticle adhesion forces being measured, while remaining low enough to allow easy removal and replacement of the attached grain. 

Below the cantilever, a monolayer of grains is fixed onto an acrylic plate mounted on a 3D micro-piezo translation stage (Micronix PPS-20), enabling vertical motion of the particle layer. The stage can be operated manually via a joystick for approach and positioning, or automatically through computer control to perform measurement cycles.

Two DMK Imaging Source cameras equipped with Leica Z16 lenses provide high-resolution side and top-view imaging ($1.5$ micron/pix)  to ensure particle alignment (Fig.~\ref{setup}b) and continuously measure cantilever deflection. LED backlighting provides high contrast for accurate particle tracking. The entire setup is mounted on an anti-vibration table to minimize mechanical noise. 

The experiment operates like an AFM, with detachment forces inferred from measurements of cantilever deflection. However,  the simplicity of our setup allows many (bottom and top) particles to be probed without changing the cantilever. This turns out to be crucial for performing relevant statistical analysis on the measured force for our coated particles.

\subsection{Characterizing the glass beam cantilever}

One strength of the experimental setup is its ability to measure a wide range of forces by appropriately selecting the cantilever geometry. The accessible force range is determined by the cantilever stiffness and the camera field of view. In practice, these parameters are adjusted so that the maximum cantilever deflection at detachment remains within the camera field of view, usually set with a height between 1 and 2 mm. 

The cantilever stiffness $k$ is given by beam theory~\cite{Truesdell1960}:
\begin{equation}
k=E\frac{we^3}{4 L^3} \,,
\label{eq:raideur}
\end{equation}
where $e$ is the beam thickness, $L$ its length, $w$ its width and $E$ its Young's modulus. 

To characterize the accessible stiffness range, we varied the beam thickness $e$ from 145 to 400 $\mu$m and the length $L$ from 6 to 8 cm, while keeping the width constant at $w = 1$ mm. We measured the force as a function of deflection using a precision scale. The results, shown in Fig.~\ref{setup}c, demonstrate that the stiffness can be varied over two orders of magnitude, ranging from 0.10 to 14.25 N.$\mbox{m}^{-1}$, following the predicted law given by eq.\ref{eq:raideur}. We also verified that different glass beams with the same geometry exhibit the same stiffness (Fig.~\ref{setup}c).

Unless otherwise specified, a glass beam with $L = 8$ cm, $e = 0.145$ mm, and $w=1$ mm was used for most of the adhesion force measurements. The camera field of view sets the maximum measurable deflection, corresponding to a maximum force of 200 $\mu$N. With a camera resolution of $1.5$ micron/pix, the minimum detectable force is approximately 0.2 $\mu$N. The setup therefore allows force measurements over a range spanning approximately three orders of magnitude, from 0.2 to 200 $\mu$N (see \ref{adaptability_appendix}).

\begin{figure*}[t!]
  \centering
  \includegraphics[width=\textwidth ]{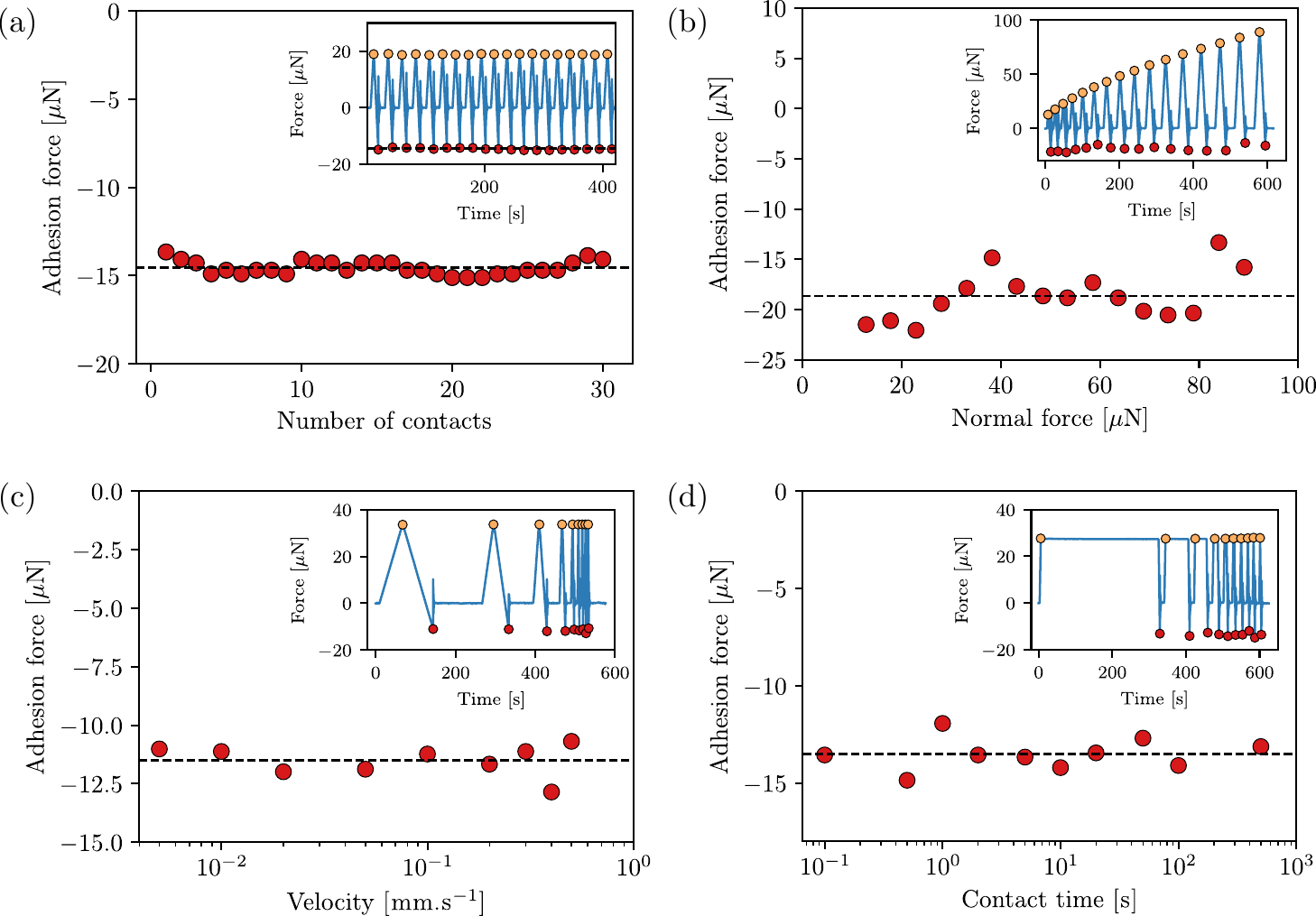}
  \caption{ Adhesion force as a function of (a) the number of repeated contacts, (b) the normal loading force prior to detachment, (c) the pulling velocity, and (d) the contact time before detachment. All measurements were performed for coated particles with $d = 199\,\mu$m and $b=65$ nm. Insets show the time evolution of the force, measured by the beam deflection, during the tests.}
  \label{aging_test}
\end{figure*}

\subsection{Experimental protocol for force measurements}

The first step of the protocol involves fixing the particles.  A small amount of repositionable adhesive is applied to the cantilever tip over a region of approximately 1 mm in length, and a thin layer of the same adhesive is sprayed onto the base plate. One particle is then attached to the free end of the cantilever, while a monolayer of particles is fixed to the acrylic base plate. The cantilever is then clamped to the vertical support, and the acrylic plate is mounted on the 3D piezo stage.

Once the particles are mounted, top -and side- view images, combined with joystick control, are used to align the particle attached to the cantilever with a particle on the substrate, as shown in Fig.~\ref{setup}b. When the two particles are vertically aligned, the measurement procedure begins. The acrylic plate is first raised, bringing the particles into contact and producing an upward deflection of the cantilever by a magnitude $\Delta z$ from its equilibrium position. This generates a normal force $F_n=k\Delta z$, which reaches its maximum value at the maximum vertical displacement of the cantilever, as depicted by the yellow circle in Fig.~\ref{setup}d.

The plate is subsequently lowered, reducing the cantilever deflection. Due to adhesive interactions between the particles, the cantilever continues moving downward with the plate beyond its equilibrium position ($\Delta z = 0$). When the cantilever reaches a critical negative displacement (red marker in Fig.~\ref{setup}d), the restoring elastic force exceeds the interparticle adhesion force $F_c$, leading to particle detachment. After detachment, the cantilever oscillates before returning to its equilibrium position. To determine the average adhesion force for a given contact, this procedure is repeated four times for each particle pair, corresponding to four successive times the cycle described in Fig.~\ref{setup}d. Measurements are performed for 50 different particle pairs, resulting in a total of 200 measurements of $F_c$ for each cohesive material.

\subsection{Sensitivity analysis}

In the protocol described above for measuring adhesion force, several experimental parameters may influence the measured force. We therefore first verified that the measured adhesion force is insensitive to the number of previous contacts between particles, the maximum normal load applied prior to detachment, the detachment velocity, and the contact time before detachment. Figure \ref{aging_test} summarizes four tests performed on CCGM particles with a diameter $d=199\,\mu$m and a coating thickness $b=65$ nm. The insets show the time evolution of the force corresponding to the different experimental protocols.

The first test, shown in Fig.~\ref{aging_test}a, indicates that the measured adhesion force remains unchanged after repeated loading cycles (30 detachments under identical loading conditions on the same contact). This suggests that repeated contacts do not significantly alter the coating properties. Figure \ref{aging_test}b further demonstrates that the adhesion force is independent of the maximum normal load applied before detachment. Similarly, by varying the pulling velocity from 0.01 to 1 mm/s (Fig.~\ref{aging_test}c) and the contact time from 0.1 to 500 s (Fig.~\ref{aging_test}d), we observe that the adhesion force remains constant. These results indicate that over the investigated parameter ranges, the adhesion force is primarily determined by the particle and coating properties.

Unless otherwise specified, all subsequent experiments were performed with a normal load of 40 $\mu$N and a detachment velocity of 20 $\mu$m/s.

\begin{figure*}[t!]
  \centering
  \includegraphics[width=\textwidth ]{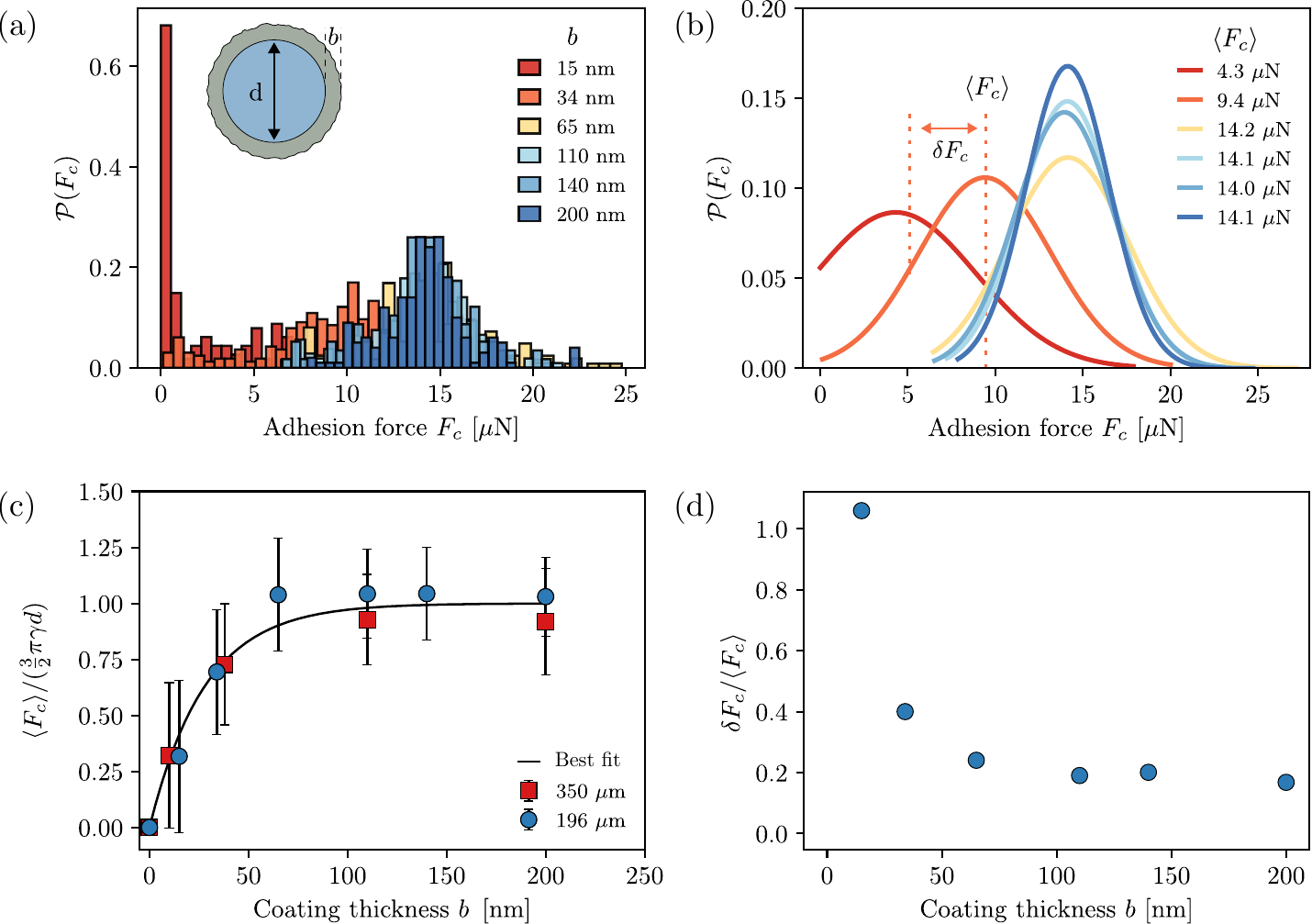}
  \caption{(a) Distribution of adhesion force ${\cal P}(F_c)$ for $d=199\, \mu$m and different coating thicknesses $b$. (b) Corresponding Gaussian fits for each $b$. (c) Dimensionless average adhesion force as a function of $b$ for $d = 199\,\mu$m (blue circles) and $350\,\mu$m (red squares); the solid line corresponds to the fit given by  Eq.~\eqref{eq:force_x_b} with $\gamma=16$ mN/m and $B = 80$ nm. (d) Relative width of the distribution, $\delta F_c/\langle F_c\rangle$, as a function of $b$.}
   \label{force_result}
\end{figure*}

\subsection{Results for the adhesion force}

Figure \ref{force_result} presents adhesion force measurements performed on CCGM particles with coating thicknesses $b=15,34,65,100$ and $200$ nm and particle diameters $d=199$ and $350$ $\mu$m. For each particle batch, measurements were performed on 50 pairs of particles, with 4 measurements per pair. Figure \ref{force_result}(a) shows the resulting force distributions.

Except for the thinnest coating ($b=15$ nm), for which a significant fraction of particle pairs exhibited no measurable adhesion, resulting in a peak at zero force, the measurements display approximately Gaussian force distributions. For each coating thickness, we fitted the distributions using a Gaussian function to extract the mean adhesion force $\langle F_c(b)\rangle$  and the distribution width $\delta F_c(b)$, which reflects heterogeneity in the coating thickness. Note that for $b=15$ nm, we retained the same definition based on the Gaussian fit, although the distribution is not Gaussian. We verified that it affects the average force and distribution width by less than 10\%.

As shown in Fig.~\ref{force_result}b, increasing $b$ both narrows the distributions and shifts them toward larger adhesion forces. Figure \ref{force_result}c shows the average adhesion force $\langle F_c(b)\rangle$ as a function of coating thickness $b$. The force increases with increasing $b$ before reaching a plateau at large coating thickness. To compare the two particle sizes, we non-dimensionalize the data using the capillary force scale $3\pi \gamma d/2$, where the surface tension $\gamma$  is chosen such that the normalized force approaches unity at large $b$. The results for the two particle sizes collapse onto a single curve, indicating that this capillary scaling captures the relevant physics. The data are well described by:
\begin{equation}
\langle F_c \rangle=\frac{3}{2}\pi \gamma d \left( 1- e^{-b/B} \right),
\label{eq:force_x_b}
\end{equation}
where  $B$ is a characteristic length scale.

This behavior is similar to observations reported in~\cite{gans2020cohesion} for larger particles ($d > 800\,\mu$m). The best fit yields $\gamma=16$ mN/m and $B \approx 28$ nm. The value of $\gamma$ is close to both the surface tension of PDMS and the value reported in~\cite{gans2020cohesion}. However, $B$ was previously found to be approximately 230 nm for larger particles. We attribute this difference to particle roughness: for our particles, we obtain an RMS roughness of approximately 25 nm (measured with AFM), which is smaller than that reported in~\cite{gans2020cohesion}.

Figure \ref{force_result}d shows the relative width of the force distribution $\delta F_c/\langle F_c \rangle$ as a function of $b$. It decreases from approximately 100\% at low coating thickness to about 15\% at large coating thicknesses. These results support the interpretation that increasing the coating thickness progressively fills surface asperities, thereby increasing the average adhesion force. Once the asperities are fully covered, particle contacts occur predominantly between polymer-coated surfaces, and further increases in coating thickness no longer increase the adhesion force.

\section{Measuring detachment energy}\label{sec:energy}

Several numerical studies of the rheology of cohesive granular materials \cite{RoyEtAl2016,mandal2020insights,mandal2021rheology,MacauleyRognon2021} suggest that understanding their flow properties requires more than just knowledge of the interparticle adhesion force. The energy required to dynamically detach particles, which accounts for both the range of adhesive interaction and dissipative processes during detachment, may also be a crucial microscopic quantity to consider. This idea is also supported by \cite{lamarche2021}, who show that an agglomeration energy controls the dilute flow regime in different configurations. This motivated us to develop a novel method to quantify the energy involved during the dynamic debonding of two particles.

In this section, we demonstrate how our cantilever setup can be adapted to operate in a \lq\lq catapult\rq\rq~mode, providing access to the detachment energy. The principle involves launching a particle attached to another at a given velocity and determining whether the initial kinetic energy is sufficient to break the adhesive bond.

\begin{figure*}[h!]
  \centering
  \includegraphics[width=\textwidth ]{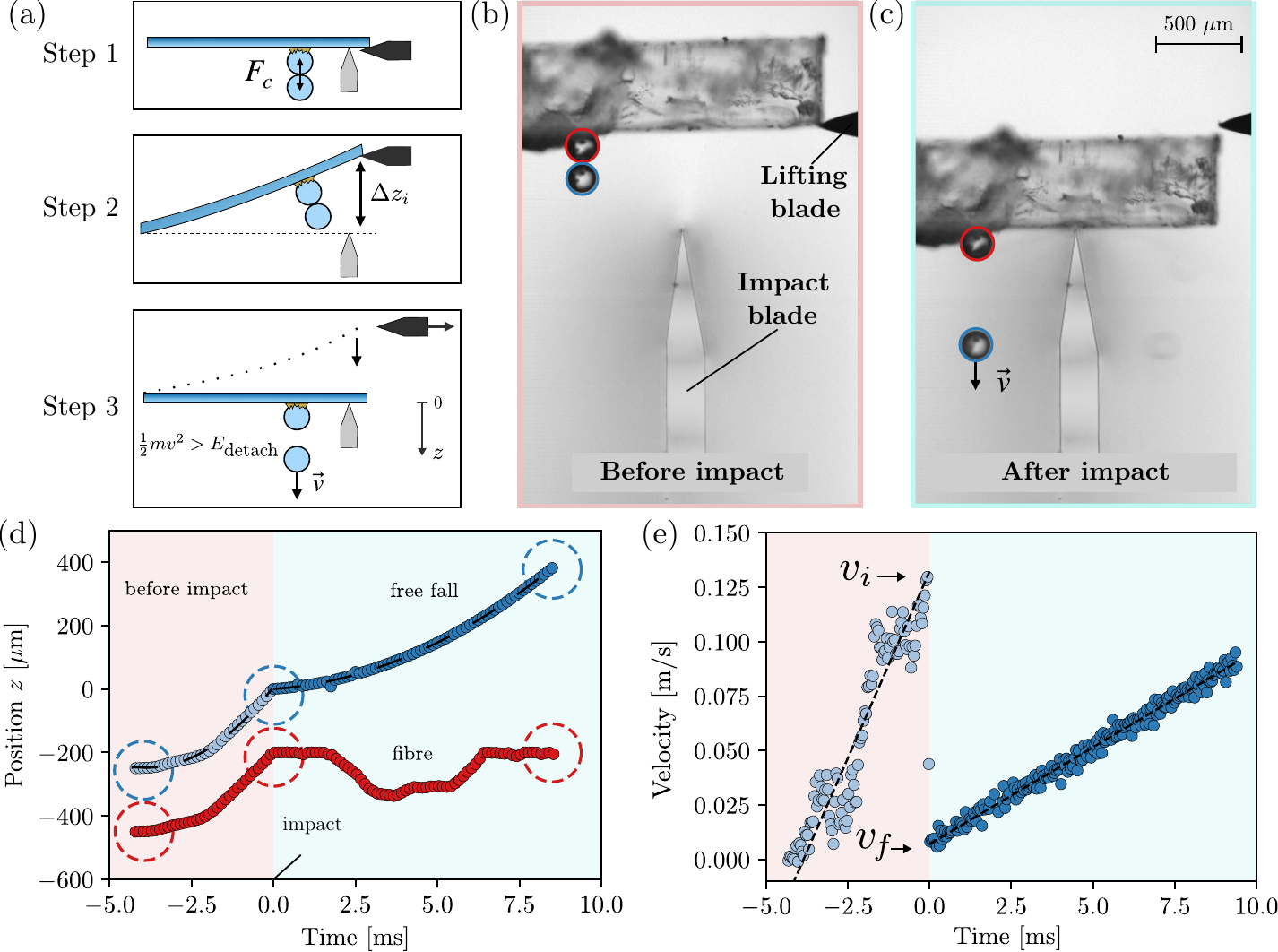}
  \caption{(a) Sketch of the catapult experiment; Step 1: a pair of particles is attached to the cantilever; Step 2: the beam is bent to an initial vertical deflection $\Delta z_i$; Step 3: the beam is suddenly released and impacts a fixed blade. (b, c) Images of the setup corresponding to Step 2 (prior to impact) and Step 3 (after impact). (d) Positions of the launched particle (blue) and the fixed particle (red) as a function of time;  $t=0$ corresponds to the time of impact. (e) Velocity of the launched particle as a function of time (derivative of data from (d)). }
    \label{catapult}
\end{figure*}

\subsection{From a cantilever to a controlled catapult}

The measurement principle and protocol used to determine the detachment energy are illustrated in Fig.~\ref{catapult}. First, one particle is attached to the cantilever tip, while a second free particle is brought into contact with it and held in place solely by adhesion forces (Step 1). The cantilever is then bent upward to an initial displacement $\Delta z_{i}$ using a horizontal razor blade mounted on a piezo actuator (Step 2). The blade is subsequently retracted, allowing the cantilever to accelerate downward until it impacts a vertical razor blade, which abruptly stops its motion (Step 3). If the initial deflection $\Delta z_{i}$ is sufficiently large, the kinetic energy of the particle immediately before impact exceeds the adhesion energy binding the particles together, causing the lower particle to detach. The kinetic detachment energy $E_{\mbox{detach}}$ is defined as the minimum kinetic energy required for particle detachment. 
To achieve sufficiently high velocities capable of overcoming the adhesive interactions investigated here, a cantilever stiffer than the one used for force measurements was used: a glass beam with a thickness of 550 $\mu$m and a length of approximately 6 cm.

To accurately characterize the particle dynamics, the side camera was replaced with a high-speed camera (Phantom v7) equipped with a Z16 LEICA lens, recording the detachment process at 24,000 frames per second. The positions of both particles were tracked using a Hough-transform-based algorithm~\cite{yang2014fast}, with sub-pixel accuracy achieved through a Gaussian fit to detect the peak intensity at the center of the grains (Fig.~\ref{catapult}c).

Typical tracked trajectories are shown in Fig.~\ref{catapult}d. The red curve represents the position of the particle attached to the cantilever, which accelerates and then rebounds after impact at $t=0$. The blue curve corresponds to the free particle launched after the impact. The first phase ($t<0$, pink zone) corresponds to the acceleration phase prior to impact, during which the particles remain attached. The second phase (blue region) corresponds to the post-detachment trajectory, where the launched particle follows free-fall motion. The launch velocity $v_i$ is controlled by the initial cantilever deflection $\Delta z_{i}$.

The detachment process is associated with an abrupt change in velocity, as shown in Fig.~\ref{catapult}e, where the particle velocity (obtained by differentiating its position) is plotted as a function of time. The velocity of the particle, initially attached to the cantilever, first increases linearly during the beam acceleration, then undergoes a sudden drop during impact, and subsequently evolves under gravity.

The detachment process itself is very rapid and occurs on length and time scales not captured by our measurements. However, from the data, we can determine the velocity just before impact, $v_i$, and the final velocity just after detachment,  $v_f$, by fitting the trajectories with a uniformly accelerated equation of motion, as depicted by the dashed lines in Fig.~\ref{catapult}d and \ref{catapult}e.

\begin{figure}[h!]
  \centering 
  \includegraphics[width=0.9\columnwidth ]{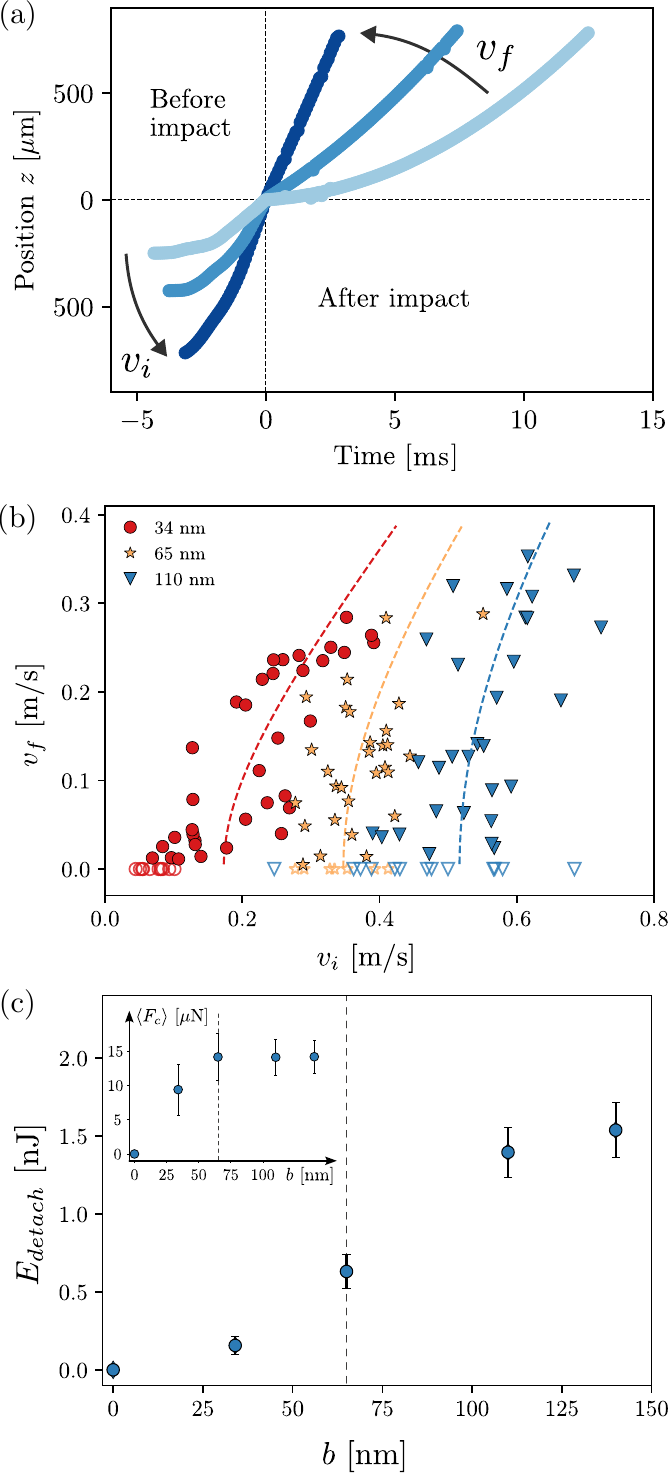}
  \caption{(a) Position of the launched particle vs time for an increasing initial bending of the cantilever from light to dark blue. (b) Velocity just after impact, $v_f$, as a function of the pre-impact velocity $v_i$. Dashed lines represent fits using the square root function $\displaystyle v_f=\sqrt{v_i^2-v_{0}^2}$. c) Kinetic detachment energy $E_{detach}=\frac{1}{2}m v_{0}^2$ as a function of the coating thickness $b$ for $d = 199$ $\mu$m;  inset: average adhesion force $\langle F_c \rangle$ for comparison.}
   \label{vf_vi_plots}
\end{figure}

\subsection{Results for the kinetic detachment energy}

Here, we present energy measurements for CCGM particles with a diameter  $d = 199$ $\mu$m and coating thicknesses $b=34$, $65$, $110$, and $140$ nm. For each value of $b$, more than 50 catapult experiments were conducted, varying the impact velocity $v_i$ by adjusting the initial deflection of the cantilever. Figure \ref{vf_vi_plots}a displays the trajectories of the launched particle for varying initial cantilever bending. Both the pre-impact velocity $v_i$ and post-impact velocity $v_f$ increase with greater initial bending.

Figure \ref{vf_vi_plots}b shows how the post-impact velocity $v_f$ varies with the impact velocity $v_i$. The different symbols correspond to different coating thicknesses. The filled symbols represent experiments where the grains detached, while the empty symbols represent experiments where the initial kinetic energy was insufficient and the grains remained attached. Although there is a relatively wide dispersion in the data, likely due to the heterogeneity of the coating, we systematically observe that the post-impact velocity becomes non-zero at a finite impact velocity and, on average, increases above this critical value.

Close to the detachment threshold, when $v_f$ is small, the energy dissipated during detachment may be considered constant. Hence, to determine the minimal impact velocity $v_0$ required for detachment,  we fit the data with a square root function $\displaystyle v_f=\sqrt{v_i^2-v_{0}^2}$ (dashed line in Fig.~\ref{vf_vi_plots}b). By slightly varying the range over which the fit is made, we estimate an uncertainty for each type of particle, ranging from $\Delta v_c = 0.05 \, \rm{m.s^{-1}} $ for the least coated particles to $\Delta v_c = 0.1 \, \, \rm{m.s^{-1}}$  for the most coated. The kinetic detachment energy is then estimated as  $E_{\mbox{detach}}=\frac{1}{2}m v_{0}^2 $. 

Figure \ref{vf_vi_plots}c shows how the kinetic detachment energy varies with the coating thickness $b$. In contrast to the adhesion force, which saturates for $b \gtrsim  60$ nm (inset of Fig.~\ref{vf_vi_plots}c), the kinetic detachment energy continuously increases with $b$ for all values investigated here. These results demonstrate that force and energy are distinct quantities, as energy is highly sensitive to dissipation mechanisms. Characterizing both is essential for a comprehensive understanding of the adhesion properties of cohesive powders.
\begin{figure}[h!]
  \centering
  \includegraphics[width=0.9\columnwidth]{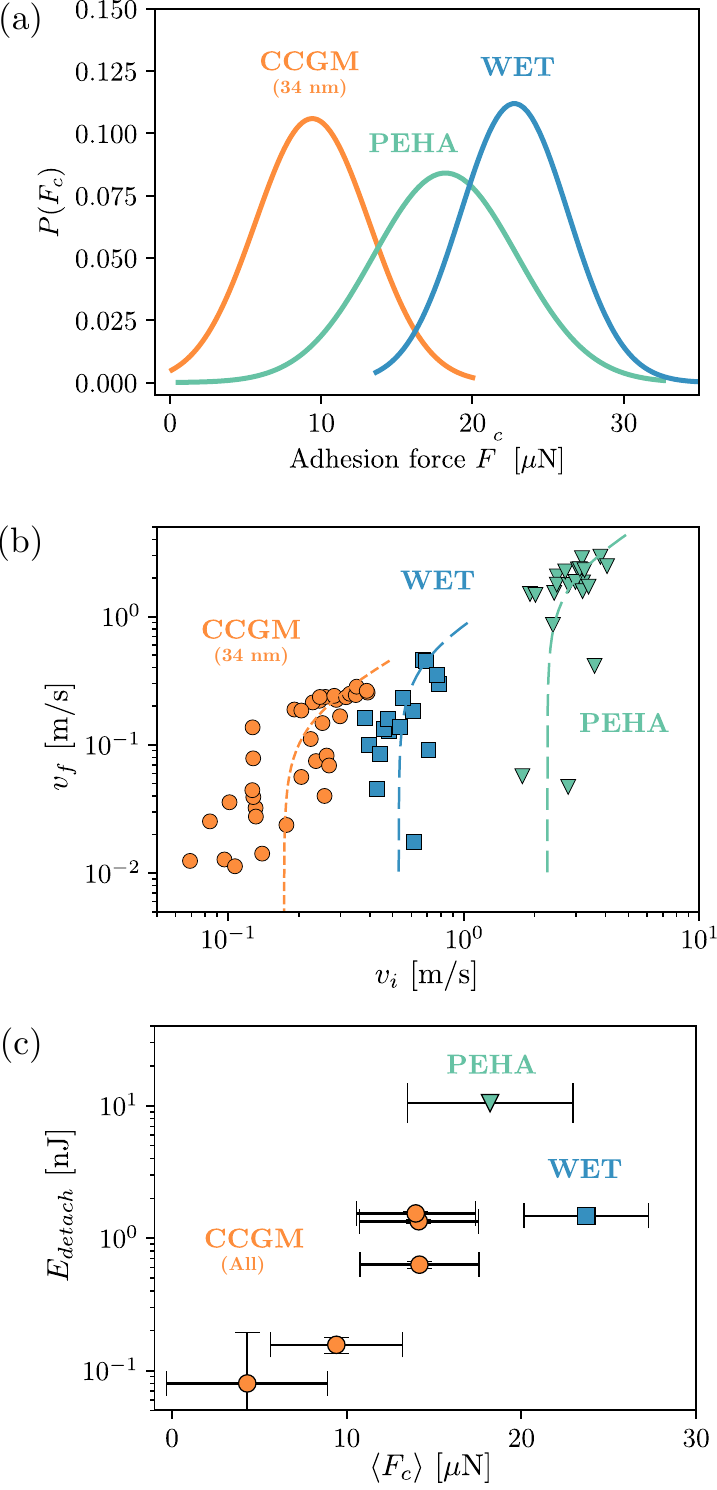}
  \caption{Comparison between the three types of particles: CCGM particles ($b = 34$ nm), wet particles and PEHA particles. (a) Gaussian fits of the adhesion force distributions.(b) Final velocity after being catapulted as a function of the impact velocity. (c) Kinetic detachment energy as a function of the adhesion force for all the particles tested in this study. The orange dots correspond to different CCGM coating thicknesses $b$ (15, 34, 65, 110 and 140 nm, from bottom to top). Error bars, when not visible, are smaller than the symbol size.}  \label{comparison_1}
\end{figure}

\subsection{Origin of the kinetic detachment energy} 

We define the kinetic detachment energy as the minimum kinetic energy required to separate one particle from another. While this quantity is experimentally well-defined, its interpretation is more complex. It is important to emphasize that $E_{detach}$ does not correspond to the work of adhesion, which is given by the integral of the force over the interaction range. Instead, it also includes the dissipation occurring during the dynamic debonding of the particles. For coated particles, this dissipation arises from the stretching and breaking of the polymer film at the interface between the two particles. Hence, the kinetic detachment energy $E_{detach}$ captures the effects of both conservative forces (adhesion) and non-conservative forces (viscoelastic forces and plastic deformation). Such a quantity is not common in the literature, although it might be related to the concept of agglomeration energy introduced in \cite{lamarche2021}, which corresponds to the minimum kinetic energy required for particles to avoid being trapped during a collision. The authors have shown that this quantity, which is also sensitive to dissipation processes, controls the dilute flowing regime of diverse granular flows. The kinetic detachment energy, being easier to measure, may play a similar role.

To quantify the relative importance of dissipation in $E_{detach}$, we can estimate the work of the adhesion force when detaching two coated particles. Using the values obtained experimentally in Section~\ref{sec:force}, the order of magnitude for the typical work of adhesion, $F_c \times b $, is approximately $10^{-12}$ J, which is about 1000 times smaller than the typical kinetic detachment energy we measured. This suggests that the detachment process in the CCGM particles is primarily controlled by dissipation rather than the work of adhesion.

To better understand the dynamics of detachment, we present in \ref{appendix2} a theoretical model for the separation of two particles connected by a thin viscoelastic bridge. In this model, the fluid is treated as a Maxwell liquid, and the geometry of the bridge is simplified to a cylinder connecting the two particles. An empirical model is introduced to describe the transition of the liquid elongation from a thin layer (governed by lubrication equations) to an elongated filament before rupture.

Under these assumptions, the model captures the existence of two regimes, depending on the initial relative velocity of the particles. At low velocities, the launched particle initially moves away but is pulled back and remains attached to the other grain. At high velocities, the particle detaches and moves away indefinitely from the other grain. Using rheological properties from the literature for the PBS coating and assuming that the initial volume of the capillary bridge is equal to the total volume of coating around the particles, the model reproduces the order of magnitude of the critical detachment velocity measured in the experiments. However, it remains unclear whether, in our experiments, detachment occurs through the elongation of a viscoelastic bridge or by fracture, which would require a different modeling approach beyond the scope of this paper.

\section{Comparison Between Different Cohesive Materials}\label{sec:DiffMaterials}

In this final section, we apply the cantilever and catapult methods to measure the adhesion force distribution as well as the kinetic detachment energy for two other cohesive materials introduced in Section~\ref{sec:particles}: wet granular materials and cross-linked PEHA polymer particles.

Figure \ref{comparison_1}a compares the force distributions for wet grains and PEHA particles with CCGM particles ($b=34$ nm). For all three materials, the adhesion force is of the same order of magnitude, with the wet particles being the most adhesive, followed by the PEHA particles, which are slightly more adhesive than the CCGM coated particles. The shift between CCGM and wet particles could come from a difference in surface tension, which is higher for the glycerol mixture ($60$ mN.$\mbox{m}^{-1}$) than for PDMS ($20$ mN.$\mbox{m}^{-1}$). The origin of adhesion for the PEHA particles is more complex to interpret. It may result from the deformability of the rubbery microparticles and the intricate interactions between the dangling chains from the polymer network at the surface of the synthesized microparticles.
 
In Fig.~\ref{comparison_1}b, we present the results of the catapult experiments, comparing how the post-impact velocity varies with the initial velocity for the three materials. In contrast to the adhesion force, the kinetic detachment energies measured for the different materials differ by more than one order of magnitude. The PEHA particles require a significantly higher initial kinetic energy to detach compared to the wet grains, which in turn require more energy to detach than the 34 nm coated CCGM particles. Figure \ref{comparison_1}c summarizes these findings in a force-energy diagram for the materials tested in this study, demonstrating that these materials cover a relatively wide range of parameters.

\section{Conclusion} \label{sec:conclusion}

This study presents an innovative cantilever-based method to characterize interparticle adhesion in cohesive granular materials. Importantly, the method bridges a critical gap by enabling adhesion measurements for particle sizes in the range of 100 $\mu$m to a few millimetres (see \ref{adaptability_appendix}), where standard AFM-based methods are hardly applicable. Additionally, it allows for measurements across a large number of particle pairs, which is crucial for obtaining statistically relevant data. By combining a glass-beam cantilever for force measurements with a novel catapult mode for energy measurements, the method provides access to two complementary adhesive properties: the quasi-static detachment force and the kinetic detachment energy.

To demonstrate the range of applicability of the method, we applied it to three distinct cohesive materials: polymer-coated particles (CCGM), wet granular media, and low $T_g$ cross-linked polymer microparticles. For CCGM particles, the adhesion force follows a capillary scaling with coating thickness and saturates beyond a characteristic coating thickness, consistently across two particle sizes. By contrast, the kinetic detachment energy continues to grow with coating thickness well beyond this saturation point, highlighting the fundamental decoupling between force and energy as complementary adhesive properties.  Across the three materials tested, the adhesion force varies by a factor 5, whereas the kinetic detachment energy spans over two orders of magnitude, demonstrating that these model materials cover a wide range of adhesive parameters, which could be of interest for future studies investigating their bulk properties.

Fundamentally, this method opens new avenues for connecting microscopic adhesive properties to macroscopic flow behavior in cohesive granular media. The ability to independently measure the quasi-static detachment force and kinetic detachment energy will be crucial for identifying the role of each interparticle adhesive property in the different regimes of cohesive rheology. This may be the key to shedding light on the physical origin of powder flowability, ultimately advancing both fundamental understanding and practical applications in industries where cohesive powders play a central role.

\section*{CRediT Author Statement}
\begin{itemize}
\item[-] M. Lajeunesse: Methodology, Investigation, Data Curation, Formal Analysis, Writing-Original Draft. 
\item[-]F. M. Rocha: Conceptualization, Supervision, Writing-Review \& Editing. 
\item[-]D. Dumont: Methodology, Formal Analysis, Writing-Review \& Editing. 
\item[-]V. Bertin: Methodology, Formal Analysis, Writing-Review \& Editing. 
\item[-]Y. Moratille: Resources. 
\item[-]N. Sintes-Zydowicz: Resources, Writing-Review \& Editing. 
\item[-]E. Drockenmuller: Resources, Writing-Review \& Editing. 
\item[-]O. Pouliquen: Conceptualization, Funding Acquisition, Supervision, Writing-Review \& Editing.
\end{itemize}

\section*{Acknowledgments}
This work greatly benefited from discussions with M. Nicolas and A. Asnacios, who advised us on the cantilever setup. We also thank A. Fruh, B. Bresson, and J. Comtet for their help in performing a few AFM experiments. This project has received funding from the European Research Council (ERC) under the European Union's Horizon Europe programme (Grant No. 101097842). The authors also acknowledge financial support from the International Fine Particle Research Institute (IFPRI) and the Agence Nationale de la Recherche under the RheoCom project (Grant No. ANR-22-CE06-0020). Views and opinions expressed are, however, those of the author(s) only and do not necessarily reflect those of the European Union or the European Research Council. Neither the European Union nor the granting authority can be held responsible for them.

\bibliography{biblio_powdertech}

\clearpage
\appendix

 \section{Coating of CCGM particles}
\label{appendix0}

 We summarize here the protocol of~\cite{gans2020cohesion},  to prepare the coated silica particles (CCGM) characterized in the adhesion experiments.\\
 
\begin{figure}[h!]
  \centering
  \includegraphics[width=0.8\columnwidth ]{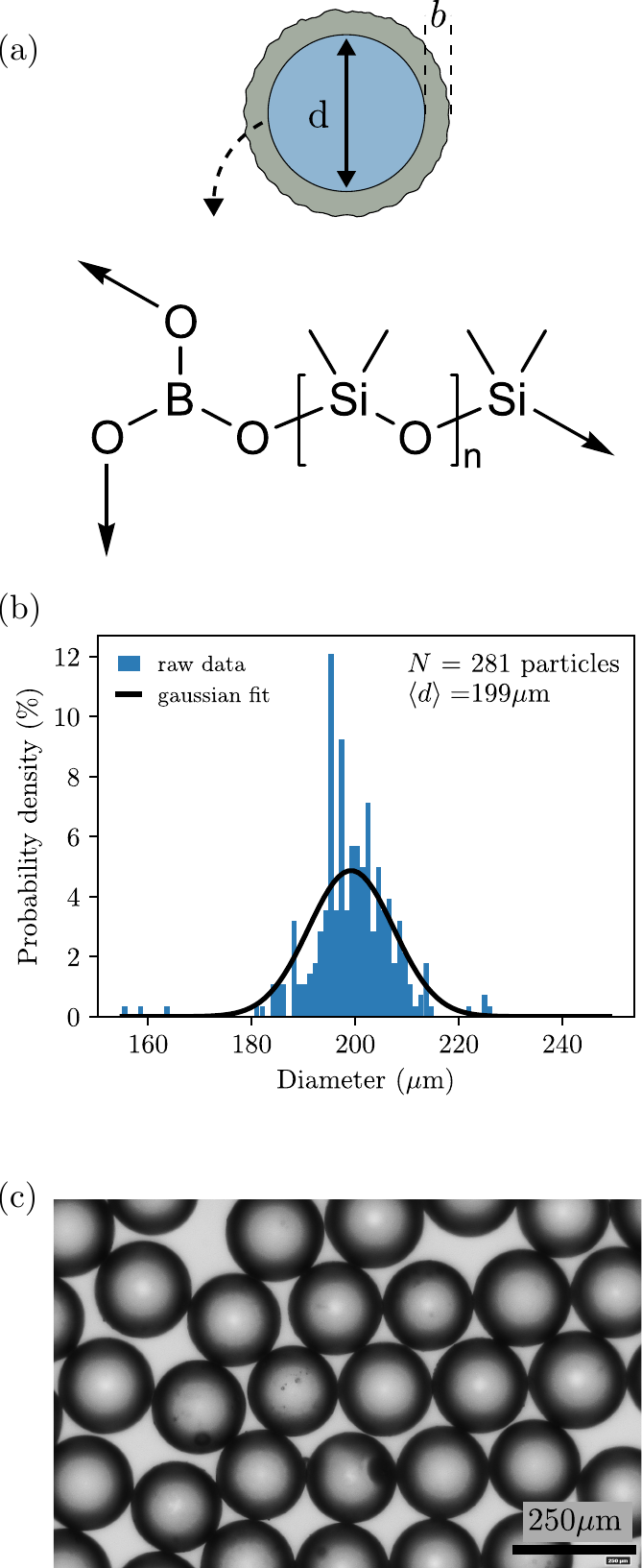}
  \caption{(a) Chemical  structure of  cross-linked polyborosiloxane (PBS). (b) Size distribution of our particles. (c) Image of CCGM particles ($b =$ 110 nm).}
  \label{CCGM_synthesis}
\end{figure}

 In this process, we used silica particles with a mean diameter of $d = 199 \pm 16~\mu$m as shown by the distribution in Fig.~\ref{CCGM_synthesis}b. First, 1 kg of dry silica particles is weighed in a large metal bowl. A sample of 50–100 g of particles is then transferred to a small glass container. Using a precision balance, a predetermined mass $m_{pdms}$ of hydroxy-terminated PDMS (with viscosity 750 cSt, Fig.~\ref{CCGM_synthesis}a) is introduced into the glass container using a syringe pump. The PDMS is poured at the center of the container to prevent contact between the liquid and the container walls, thereby minimizing material loss. The mass of PDMS is calculated to produce a target coating thickness $b$ over the entire 1 kg batch of silica particles, assuming a homogeneous coating layer.

The particles containing the PDMS are then transferred back into the large metal bowl and mixed with the remaining silica particles. To cross-link the PDMS, a boric acid solution is prepared by dissolving a mass of boric acid powder corresponding to $m_{boric\,acid} = 0.14 \times m_{pdms}$ in 100 mL of purified water at 60°C ~\cite{gans2020cohesion}. The solution is stirred for 5 min until homogeneous and is subsequently added to the metal bowl containing the silica particles and PDMS.

The mixture is then stirred for 90 minutes in a mixer maintained at 110°C to promote cross-linking of the polymer while simultaneously evaporating the water. At the end of the process, the PDMS is covalently cross-linked into polyborosiloxane (PBS) depicted in Fig.~\ref{CCGM_synthesis}a, hence forming a quasi-homogeneous coating around the silica particles.

\clearpage
 \section{Synthesis of PEHA microparticles}
\label{appendix1}
 
 We present here an extension of the protocol developed in  ~\cite{MoratileEtAl2022} using membrane emulsification-assisted suspension radical copolymerization, to prepare the soft polymer particles characterized in the adhesion experiments.\\
 
\begin{figure}[h!]
  \centering
  \includegraphics[width=0.7\columnwidth ]{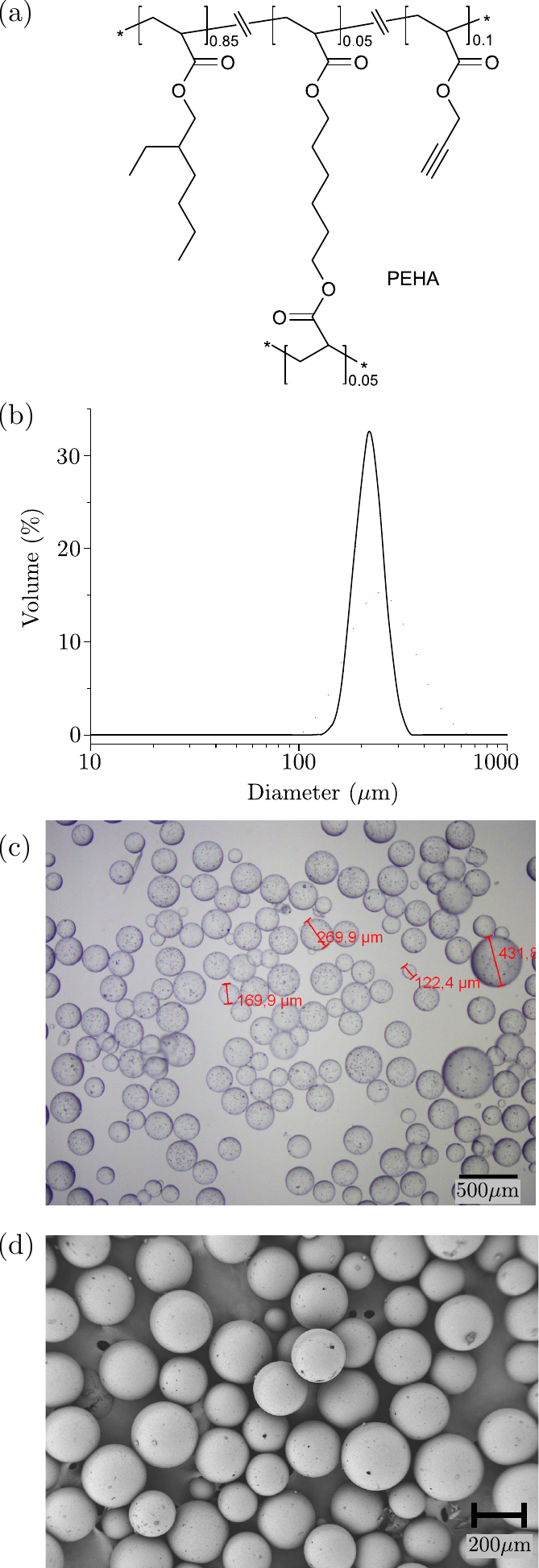}
  \caption{(a) Chemical structure, (b) Light diffraction size distribution ($d=$ 232 $\mu$m, s = 0.63), (c) Optical microscopy, and (d) Scanning electron microscopy of PEHA microparticles.}
  \label{PEHA_synthesis}
\end{figure}

The cross-linked PEHA (Fig.~\ref{PEHA_synthesis}a) microparticles were prepared by radical copolymerization in suspension in a double-walled glass reactor equipped with a 6-blade stirrer, a condenser and a thermometer (Fig.~\ref{PEHA_synthesis}c). The emulsion was formed by membrane emulsification (Dispersion Cell, Micropore Technology Ltd) using a hydrophilic metal membrane with pores 100 $\mu$m in diameter, an injection base, an emulsification cell, and a two-blade stirring module equipped with a PTFE vortex breaker and powered by a 24 V motor (DC INSTEK Model PR 3060). The injection of the monomer phase through the membrane was performed using a syringe pump (KDS100 Legacy, Fisher Scientific). The continuous aqueous phase containing poly(vinyl alcohol) (PVA, Mw = 70 kg.mol$^{-1}$, 88\% hydrolyzed, 0.32 g, 4.0 g.L$^{-1}$) dissolved in distilled water (80 mL) and the dispersed monomer phase containing a mixture of 2-ethylhexyl acrylate (15.7 g, 85.3 mmol), 1,6-hexanediol dimethacrylate (1.14 g, 5.02 mmol), propargyl acrylate (1.11 g, 10.0 mmol), and lauroyl peroxide (0.40 g, 1.01 mmol) were degassed separately by argon bubbling for 15 minutes. The monomer phase was then injected into the aqueous phase through the metal membrane at a flow rate of 2 mL.min$^{-1}$ and a constant stirring of 800 rpm. The emulsion was then transferred from the membrane emulsification device to the reactor, and the copolymerization was carried out for 5 h at 60 °C under an argon flow and a stirring of 200 rpm. The reaction mixture was then cooled down, and the microparticles were isolated through several cycles of centrifugation (10 000 rpm) and dispersion in ethanol (3 × 100 mL) and distilled water (5 × 100 mL) to remove residual traces of comonomers and the PVA adsorbed onto the surface of the microparticles. Finally, PEHA microparticles were freeze-dried and recovered as a compact white powder (15.1 g, 82.2\%) (see Fig.~\ref{PEHA_synthesis}c-d).\\

 \section{A device that can be adapted to different particle sizes and force ranges}
\label{adaptability_appendix}

\begin{figure}[t!]
  \centering
  \includegraphics[width=\columnwidth ]{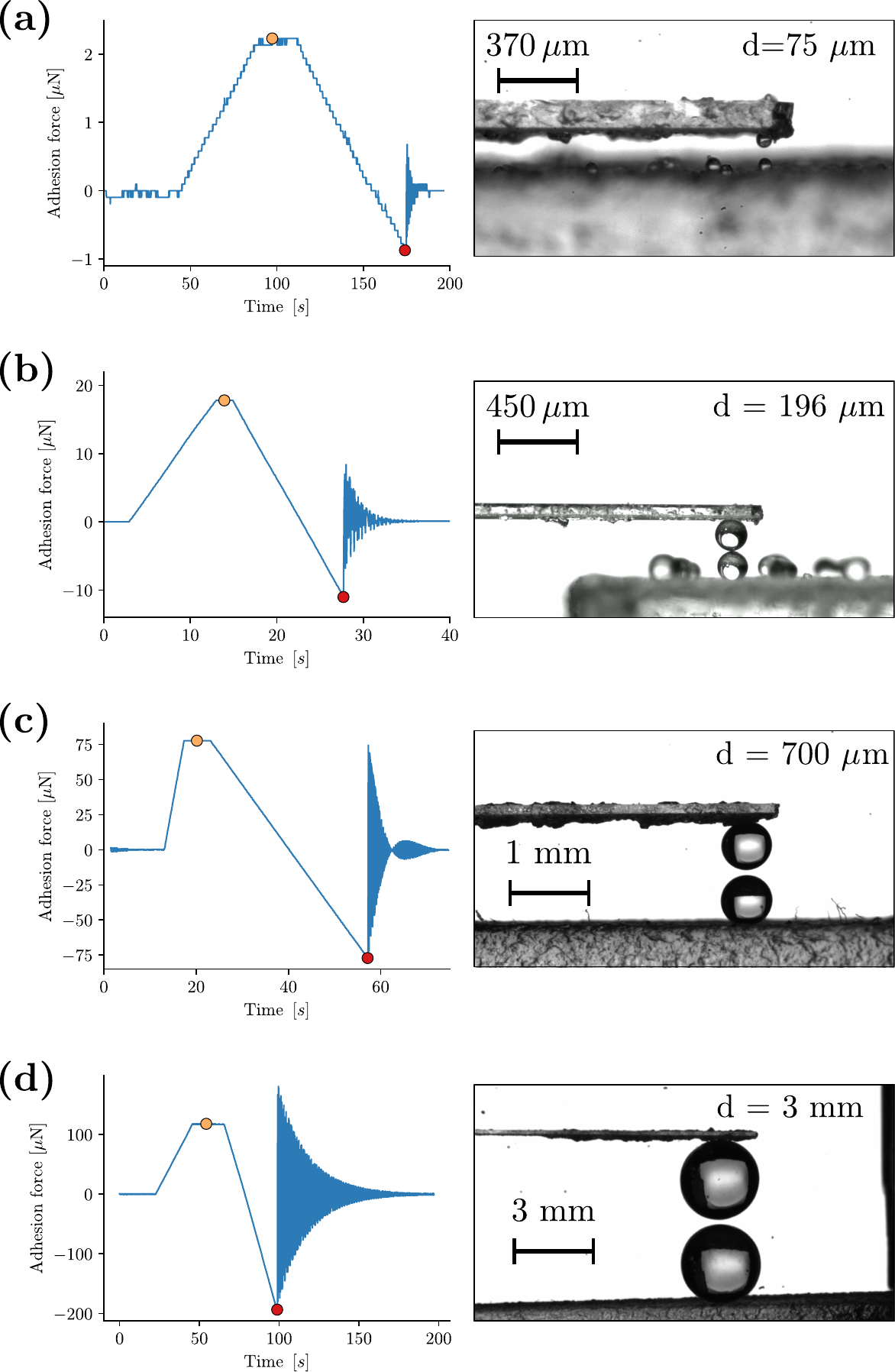}
  \caption{Measurements of adhesion force for different particle sizes using a $145 \,\mu$m thick glass fiber. (a) CCGM with $d = 75$ $ \mu$m  and $b=34$ nm. (b) CCGM with $d = 199$ $ \mu$m  and $b=65$ nm. (c) CCGM with $d = 700$ $ \mu$m  and $b = 112$ nm. (d) CCGM with $d = 3000$ $\mu$m  and $b =147$ nm. }
  \label{adaptability_fig}
\end{figure}

We show in this appendix that the cantilever experiment can be easily adapted for measuring adhesion forces across a wide range of particle sizes, from 75 $\mu$m to 3 mm. Figure \ref{adaptability_fig} presents adhesion measurements for four different particle sizes, using the same fiber. For the smallest particles (d = 75 $\mu$m, Fig~\ref{adaptability_fig}a), we measured forces in the range of 0.5-1 $\mu$N. For the largest particles (d = 3 mm, Fig.~\ref{adaptability_fig}d), the measured forces reached approximately  200 $\mu$N, highlighting the method's broad applicability. Notice that the sensitivity can be even lower when using thinner fibers. In practice, a limitation arises from the optical system, which must be able to zoom sufficiently to accurately measure the fiber deflection. We adapt the zoom for each type of particle, to keep the largest deflection inside the camera range.

\clearpage

\section{Detachment of two particles linked by a viscoelastic bond}
\label{appendix2}

In this appendix, we propose a model to estimate the kinetic detachment energy of two CCGM particles. In particular, we aim to rationalize the observed dependence of $E_{\mathrm{detach}}$ on the coating thickness (see Fig.~\ref{vf_vi_plots}b-c), whereas the quasi-static adhesion force becomes nearly independent of the coating thickness for sufficiently thick coatings (see Fig.~\ref{force_result}a-c). To that end, we develop a minimal model accounting for viscoelastic dissipation during the detachment process.

\subsection{Model geometry and momentum balance}

\begin{figure}[h!]
  \centering
  \includegraphics[width=0.5\textwidth ]{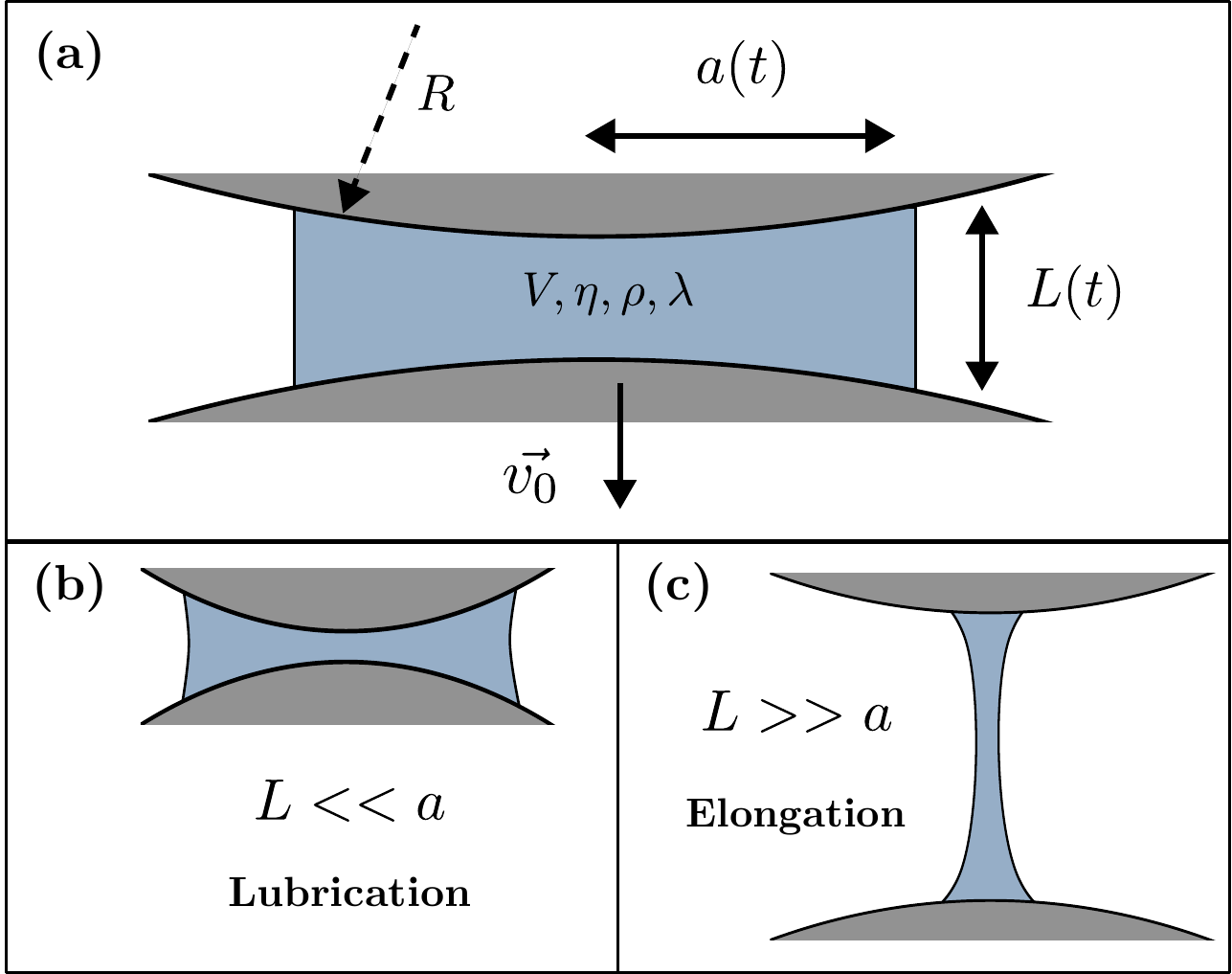}
  \caption{Schematic of a simple capillary bridge between two rigid spheres.}
  \label{viscoscheme}
\end{figure}

We consider two spheres of radius $R$ and mass $m$, linked by a capillary bridge of height $L(t)$ and wetting radius $a(t)$ (Fig.~\ref{viscoscheme}a). The total volume of fluid in the bridge is denoted by $V$. When the two particles come into contact, the (negative) capillary pressure within the connecting liquid bridge draws polymer from the coating into the contact region. In what follows, we make the simplifying assumption that the volume of the capillary bridge corresponds to the total polymer coating volume of a single particle, with each particle contributing half of its coating to the bridge:
\begin{equation}
V =  4\pi R^2 b.
\end{equation}
For the sake of simplicity, we further assume that the capillary bridge retains a cylindrical geometry throughout the separation process, such that volume conservation imposes
\begin{equation}
V = \pi a^2(t)L(t) 
\end{equation}
The liquid is assumed to behave as a viscoelastic fluid with density $\rho$, viscosity $\eta$, and relaxation time $\lambda$, while its surface tension and contact angle are denoted by $\gamma$ and $\theta$, respectively.

We consider the top particle to be fixed. The second particle is initially at a distance $L_0$ from the first and is launched with an initial velocity $v_0$. During its motion, the particle experiences two forces: a capillary force $F_{\mathrm{cap}}$ and a viscoelastic force $F_{\mathrm{viscoel}}$. The momentum balance along the vertical axis therefore reads
\begin{equation}
m\frac{d^2 L}{d t^2} = F_{cap}+F_{viscoel}.
\label{eq:EqMotion1}
\end{equation}
The following two subsections provide models for these two forces.

\subsection{Viscoelastic force}

The viscoelastic force can be expressed as the average normal stress in the fluid, $\sigma_{\mathrm{viscoel}}$, integrated over the wetted area:
\begin{equation}
\label{eq:viscoel_force}
F_{\mathrm{viscoel}}
=
\pi a^2(t)\sigma_{\mathrm{viscoel}}
=
\frac{V}{L(t)}\sigma_{\mathrm{viscoel}},
\end{equation}
where volume conservation has been used.

During detachment, the bridge geometry evolves considerably, from a nearly flat thin film (Fig.~\ref{viscoscheme}b), hereafter referred to as the \textit{lubrication limit}, to a stretched slender ligament (Fig.~\ref{viscoscheme}c), referred to as the \textit{elongation limit}. The viscous resistance originates from different flow fields in these two limiting geometries.

In the lubrication limit, $L(t)\ll a(t)$, separation of the particles induces predominantly radial flow within the thin liquid layer. Lubrication theory gives the normal viscous force
\begin{equation}
F_{\mathrm{lub}} = 6\pi\eta\left(\frac{R}{2}\right)^2 \frac{1}{L(t)}\frac{dL(t)}{dt},
\end{equation}
where $R/2$ is the reduced radius of two identical spheres. Dividing this force by the wetted area $\pi a^2(t)=V/L(t)$ gives the corresponding average stress,
\begin{equation}
\sigma_{\mathrm{lub}} = \frac{3}{2}\pi\eta R^2 \frac{1}{V} \frac{dL(t)}{dt}.
\end{equation}

In the opposite elongation limit, $L(t)\gg a(t)$, the bridge behaves as a slender filament undergoing approximately uniaxial extension. For a Newtonian fluid, the corresponding extensional stress is
\begin{equation}
\sigma_{\mathrm{elong}} = 3\eta \frac{1}{L(t)} \frac{dL(t)}{dt},
\label{eq:sigfil}
\end{equation}
where the factor $3$ is the Trouton ratio.

To describe the continuous transition between these two geometrical limits, we introduce the following interpolation for the instantaneous viscous stress:
\begin{equation}
\sigma_{\mathrm{visc}} = 3\eta \frac{1}{L(t)} \frac{dL(t)}{dt}\left[ \frac{1} {1+\displaystyle\frac{2V}{\pi R^2L(t)}} \right].
\label{eq:sigtot}
\end{equation}
This expression recovers the lubrication result for $L\ll a$ and the uniaxial elongation result for $L\gg a$. Investigating the denominator of \eqref{eq:sigtot}, the typical length at which the transition from lubrication to elongation occurs is $2V/(\pi R^2)$.

Finally, to account for the viscoelasticity of the polymer coating, we assume a linear Maxwell rheology, such that the viscoelastic stress obeys
\begin{equation}
\lambda\frac{d\sigma_{\mathrm{viscoel}}}{dt} + \sigma_{\mathrm{viscoel}} = 3\eta \frac{1}{L(t)} \frac{dL(t)}{dt} \left[\frac{1} {1+\displaystyle\frac{2V}{\pi R^2L(t)}}\right].
\label{eq:viscoel_stress}
\end{equation}

\subsection{Capillary force}

The capillary force is maximal when the two particles are close. In the limit of vanishing separation, its characteristic magnitude is
\begin{equation}
F_{\mathrm{cap}}(L\rightarrow 0)
=
2\pi\gamma R\cos\theta,
\end{equation}
which corresponds to the classical capillary force between two identical spheres connected by a liquid bridge \cite{pitois2000liquid}. As the particles separate, the capillary force progressively decreases and eventually vanishes for a sufficiently elongated bridge. For the sake of simplicity, we use the same interpolation function as for the viscous stress to connect these two limits:
\begin{equation}
 F_{\mathrm{cap}}
 =
 2\pi\gamma R\cos\theta
 \left[
 1-\frac{1}{1+\displaystyle\frac{2V}{\pi R^2L(t)}}
 \right].
 \label{eq:Fcaptot}
\end{equation}
This expression recovers the capillary force $2\pi\gamma R\cos\theta$ for $L\rightarrow 0$ and progressively vanishes as the bridge is elongated.

A similar approach was proposed by Pitois et al.~\cite{pitois2000liquid} to describe the capillary and viscous forces between two spheres connected by a Newtonian liquid bridge. Their model is based on the lubrication limit, with finite-volume corrections accounting for the finite extent of the liquid bridge. Here, we use a slightly different interpolation, as our model also aims to recover the elongational-flow limit when the bridge becomes slender, $L\gg a$. In addition, the viscoelastic nature of the polymer coating is accounted for through the Maxwell constitutive equation introduced above.

\subsection{Non-dimensionalization and numerical methods}

Equations~\eqref{eq:EqMotion1}, \eqref{eq:viscoel_force},
\eqref{eq:viscoel_stress}, and \eqref{eq:Fcaptot} form a closed
set of equations. Together with the initial conditions
$L(0)=L_0$ and $\dot{L}(0)=v_0$, they determine the dynamics
of the launched particle.

To non-dimensionalize the equations, we define
$X=L/L_c$, $t^*=t/\tau$, and $Y=\sigma/\sigma_0$, where the
characteristic length $L_c$, inertio-capillary time scale $\tau$,
and stress $\sigma_0$ are given by
 
  \begin{equation}
  \begin{aligned}
 & L_c=\frac{V}{\pi R^2}\\
 &\tau= \sqrt{\frac{m L_c}{2\pi\gamma\cos\theta R}}\\  
 &\sigma_0=\frac{m L_c^2}{V\tau^2}\\ 
 \end{aligned}
 \end{equation}
 
In dimensionless form, the system of equations and initial conditions are
\begin{equation}
\begin{aligned}
&\ddot{X}+\left( 1- \frac{1}{1+2/X^2} \right) + \frac{Y}{X}=0, \\
&\dot{Y}+ \frac{\tau}{\lambda}\left[Y-\frac{3\eta}{\sigma_0 \tau}\frac{\dot X}{X}\left(\frac{1}{1+2/X^2}\right) \right]=0, \\
&\dot{X}(0)= V_0 = \frac{v_0 \tau}{L_c},\\
&X(0)=\frac{L_0}{L_c}.
\end{aligned}
 \label{eq:system}
\end{equation}
where dots now denote derivatives with respect to the
dimensionless time $t^*$. The dynamics is therefore controlled
by four dimensionless parameters: the Deborah number
$\lambda/\tau$, a dimensionless viscosity
$3\eta/(\sigma_0\tau)$ comparing viscous and inertio-capillary
effects, analogous to the Ohnesorge number in fluid dynamics,
and lastly the dimensionless initial velocity $v_0\tau/L_c$
and initial distance $L_0/L_c$.

The system~\eqref{eq:system} is solved as an initial-value problem
using a fourth-order Runge--Kutta scheme implemented in Python
using scipy.integrate. 

\subsection{Prediction of the model}

The parameters of the viscoelastic material are taken from the data provided in~\cite{gans2020cohesion} and are fixed as $\rho_l = 970$\,kg.m$^{-3}$, $\gamma = 0.020$\,N.m$^{-1}$, and $\lambda = 3.8$\,s. The choice is less obvious for the viscosity of the cross-linked PBS coating, which is usually very high, typically between $10^3$ and $10^4$ \,Pa s, as it is strain-dependent. Here, we use $\eta = 4000$ Pa.s. The initial length of the liquid bridge is taken to be equal to the coating thickness, typically $b=110$ nm in the data presented here, which defines its dimensionless value as $X_0=b/L_c$. The total volume of the bridge is then given by $V = 4\pi R^2 b$. 

  \begin{figure*}[t!]
  \centering
  \includegraphics[width=\textwidth ]{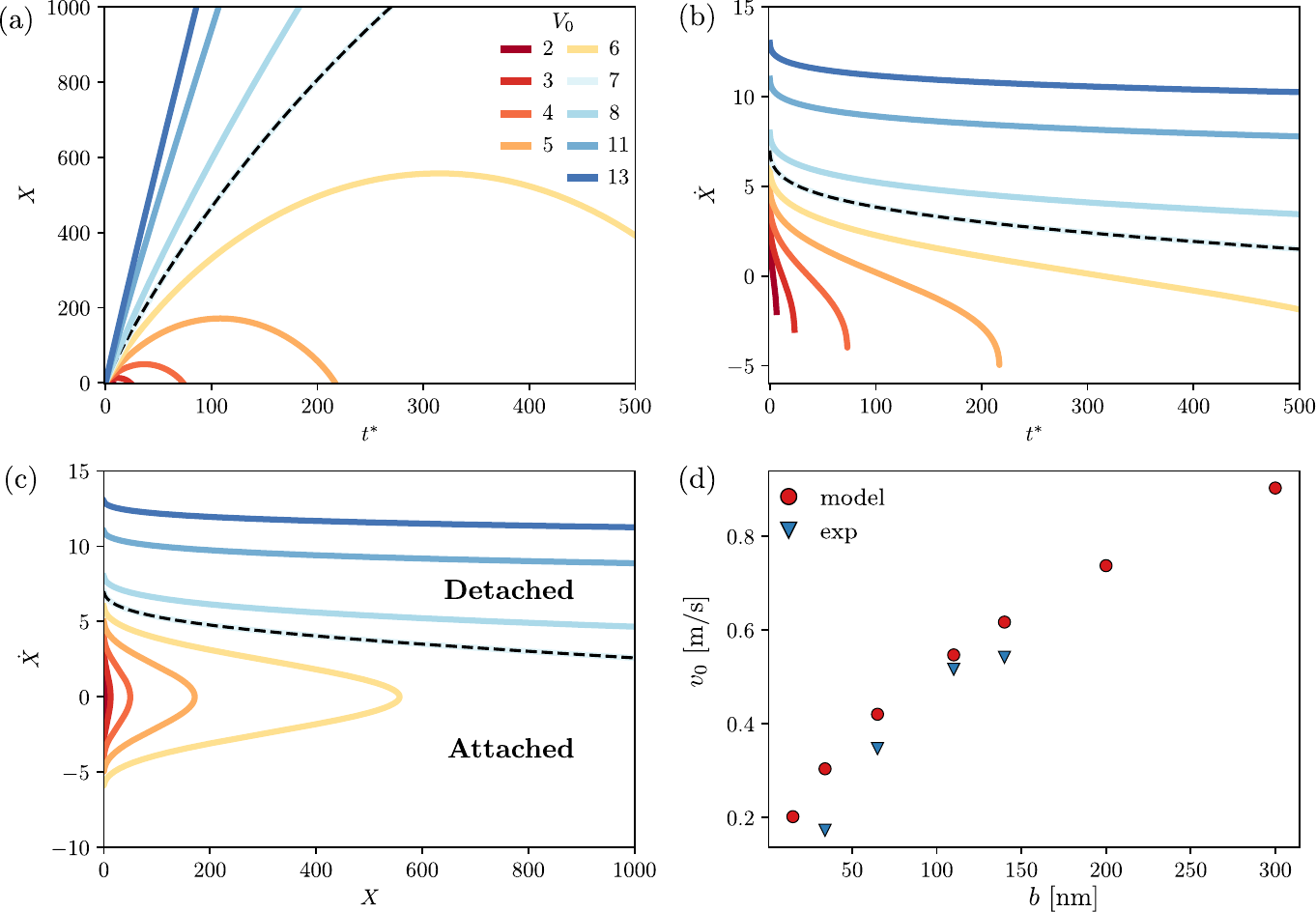}
  \caption{(a) Dimensionless position as a function of dimensionless time. The trajectories are parabolic if the grain remains attached. (b) Velocity as a function of time. The dashed black line represents the separatrix between the two regimes. (c) Phase diagram showing velocity as a function of position. (d) Escape velocities versus coating thickness, comparing the theoretical predictions of the model to the real catapult experiments.}

  \label{result_model}
\end{figure*}

The results are shown in Fig.~\ref{result_model}a--c, where each colour corresponds to a different initial velocity $v_0$. We identify two regimes: either the particle remains attached and returns to its initial position at $X = 0$, or it escapes. The transition appears clearly when plotting the velocity as a function of time, as shown in Fig.~\ref{result_model}b, or in the phase space of the trajectory, plotting $\dot X$ as a function of $X$. A velocity that tends to a positive constant indicates that the particle detaches, while a negative value indicates that the particle remains attached. Note that we stop the simulation when the particle returns to $X = 0$, as we do not model the collision. As shown in Fig.~\ref{result_model}d, the model reproduces the order of magnitude of the critical detachment velocity and qualitatively
recovers its increase with coating thickness.

\end{document}